**PUBLIC OPINION BETWEEN REARMAMENT AND CRISIS OF THE NUCLEAR TABOO**

*by Fabrizio Battistelli, Francesca Farruggia*

1. *The public's aversion for nuclear weapons between "taboos" and "tradition": what methods to study it?*

Over the past fifteen years, a significant debate has emerged among scholars concerning nuclear weapons and how they are perceived by both governments and public opinion. In parallel with the deterioration of the arms control regime between the two nuclear superpowers — evidenced by the successive abandonment of treaties regulating the possession and deployment of the arsenals of the United States and Russia[1] — the nuclear threat has gained new momentum.

In Europe in particular, the invasion of a country that had renounced nuclear weapons (Ukraine) by a country possessing 5,500 warheads (Russia) has generated growing concern among European public opinion. This has produced contradictory effects: on the one hand, renewed fears of military escalation; on the other, a resurgence of the relevance of nuclear weapons. For example, in European countries participating in the American "extended deterrence" strategy, such as Germany and the Netherlands, public opinion has shifted — from negative to positive — regarding the stationing of NATO missiles (Onderco, Smetana and Etienne 2022).

---

[1] The golden age of disarmament was the 15-year period that began in 1987 with the signing of the Treaty on the Elimination of Intermediate-Range and Shorter-Range Missiles (INF) and ended in 2002—on the one hand, with the signing of the Strategic Offensive Reductions Treaty (SORT) by the United States and Russia, and on the other, with both countries' withdrawal from the Anti-Ballistic Missile Treaty (ABMT). In 2007, Russia suspended its participation in the Treaty on Conventional Armed Forces in Europe (CFE), from which it formally withdrew in 2023. This withdrawal led to the treaty's earlier suspension and, in 2024, the official withdrawal of the United States. In rapid succession, the United Kingdom, France, Italy, Germany, Belgium, the Netherlands—as well as Belarus and Turkey—also withdrew from the CFE. In 2023, Russia additionally withdrew from the Comprehensive Nuclear-Test-Ban Treaty (CTBT). The START I treaties on strategic delivery systems expired in 2009, and SORT expired in 2011. These were replaced by the New START in 2010, which remains the last bilateral arms control agreement between Russia and the United States still in force, but it is set to expire in 2026. Furthermore, the United States withdrew from the Joint Comprehensive Plan of Action (JCPOA) on the Iranian nuclear program in 2018, and—crucially for Europe—from the INF Treaty in 2019.

These are just some of the symptoms of a progressive erosion — partly spontaneous, partly deliberate — of the phenomenon known as the "nuclear taboo." According to this concept, popularized by Nina Tannenwald (1999; 2007; 2018), the particular characteristics of atomic technology and its ultimate product, the "Bomb," have generated among the population—and to some extent among elites—a widespread sense of aversion that has so far prevented its renewed use[2].

Everything began in 1945 with the emergence on the strategic stage of a revolutionary weapon: the atomic bomb. What remains significant is the fact that, following the devastating conclusion of World War II, atomic technology has indeed proliferated to other states, but has never been used again—neither by the state that first employed it (the United States), nor by other states (primarily the Soviet Union) that adopted it in response[3].

For an history's ironies, only two nuclear warheads have been used in the atomic era, while in the first seven decades alone, approximately 125,000 were produced (Kristensen and Norris 2013). When awarded the Nobel Prize in Economics in 2005, Thomas Schelling described the non-use of nuclear weapons since the two bombs dropped on Japan as "astonishing" and referred to this fact as "the legacy of Hiroshima" (Schelling 2006) — the only positive aspect of an event universally regarded as tragic. It is worth noting, however, that following a long period of dominance by the nuclear taboo, in recent times the concept has been subject to progressive erosion, both spontaneous and deliberate. As a partial counterbalance to this trend, one can observe a renewed interest from the social sciences, particularly regarding public opinion. This has stimulated further study of the theme of "nuclear weapons use" on both a quantitative and qualitative level — on the one hand increasing the number of analyses, and on the other making them more sophisticated through the use of unconventional tools such as "experimental" surveys.

Within the lively scholarly debate on the existence and nature of the nuclear taboo, an alternative interpretation has gained ground: that the abstention from the use of such weapons is less a true taboo and more a tradition based "on the precedent and mutual adhesion models of reciprocal restraint" among international actors (Thomas 2001: 34; see also Sagan 2004; Paul 2009). Upon closer inspection, however, the definitional dichotomy between taboo and tradition is not so clear-cut — as well the debate about the application of a concept from cultural anthropology, like "taboo," appears nominalistic (Müller, 2021). In our view, the conceptual opposition between "taboo" and "tradition" does not imply an irreconcilable incompatibility. What matters is the enduring aversion — still held by a large majority — that nuclear weapons continue to provoke among the public across diverse social contexts. If we attribute to the term a heuristic rather than ontological value, it remains reasonable to use "taboo" as "shorthand for a strong normative use of nuclear weapons" (Bell 2023: 166).

---

[2] Independent and characterized by its own specificity, the so-called "renaissance" of civil nuclear power has re-emerged in several European countries—such as Italy, Germany, and the United Kingdom—which had previously appeared to abandon this technology (Renzi et al., 2017).
[3] This marks a stark contrast with what had occurred over centuries with other revolutionary military innovations, such as the horse or gunpowder, which were rapidly deployed on the battlefield in their respective eras.

Regardless of adherence to one interpretation or another, the ongoing debate has nonetheless yielded some positive outcomes. On the one hand, it has revitalized a body of contributions that radically broaden the traditional perspective of law and international relations, encouraging social scientists to invest more research into issues — such as armaments — that had long been considered "exotic" in content and difficult to verify empirically. A leading role in this regard is played by one of the branches of the social sciences most advanced in its empirical approach to phenomena: public opinion research. In order to address the question of whether or not a taboo exists in a complex domain of human behavior such as the use of a highly destructive weapon, the study of public opinion is driven to adopt innovative analytical perspectives, both conceptually and in terms of methodology and research techniques.

Before addressing conceptual aspects, we shall briefly consider the empirical ones, starting with the research associated with what has been termed "the second generation of taboo research" (Smetana and Wunderlich 2021: 1072). The goal of this new approach is to examine assessments regarding public opinion (even more so than that of elites) on key aspects of the use of force, with particular reference to nuclear weapons. One of these assessments is the frequently cited notion of the United States' specificity as a country more inclined than its allies to consider the use of weapons in international crises. In his well-known metaphor, Robert Kagan claimed that: "Americans are from Mars and Europeans are from Venus" (Kagan 2003: 3). And to those, like Johan Galtung (1996: 268), who observed that "to the man with a hammer the world looks like a nail", he replied that "nations without great military power face the opposite danger: when you don't have a hammer, you don't want anything to look like a nail" (Kagan 2003: 28).

In any case, according to the "second generation" of researchers, U.S. strategic policy has so far been studied thematically as a focus on the motivations and decisions of political elites and, methodologically, through a qualitative approach based on historical reconstructions of major events, case studies, interviews with key actors, and so on. Authors such as Sagan and Valentino propose an alternative to this approach, one that moves away from traditional opinion polls. They administer "experimental" surveys to panels of ordinary citizens, built around scenarios designed to simulate strategic emergencies as realistically as possible — contexts in which such individuals might actually have to take a position.

While we fully support the need to distinguish between the positions of decision-makers and those of ordinary citizens, we also note that the empirical study of the latter's opinions is not new. Following the events of September 11, 2001, and the subsequent "War on Terror," the use of military force prompted a significant surge of research and analysis focused on comparing American and European public opinion. For instance, regarding the 2003 war waged by President G.W. Bush against Saddam Hussein's Iraq — accused of possessing weapons of mass destruction and/or being complicit in jihadist terrorism — Pierangelo Isernia and Philip Everts, through their analysis of cross-national surveys, identified a "Transatlantic gap." Europeans differed not so much in terms of values (e.g., the democratic system and the need to defend it) from Americans, but rather in tactical preferences: placing "higher priority to soft tools" and considering the use of force as "a truly *ultima ratio* to be utilized only when all other sources of power have failed" (Isernia and Everts 2004: 255; see also Everts and Isernia 2001).

American "exceptionalism" is now being challenged by the second wave of research, particularly with regard to citizens' attitudes toward casualties suffered and inflicted in war — especially in the context of potential nuclear weapons use. The debate does not revolve around whether the American public is inclined to prioritize saving the lives of their own soldiers, even at the cost of disproportionately higher enemy casualties (combatants and non-combatants alike). In fact, "concerns about winning wars and the desire to minimize the loss of lives of their nation's soldiers dominate public opinion about military operations[4]" (Sagan and Valentino 2017: 51). Rather, the emphasis is on the fact that such a stance is by no means exclusive to the American public; it is a shared feature of Western democracies, which tend to favor the use of all available military assets (including nuclear ones) that minimize the loss of their own uniformed personnel (Sagan and Valentino 2018). Through experimental surveys, researchers have assessed the willingness of publics in other nuclear-armed countries to support the use of such weapons in specific scenarios. Using the same methods, Dill, Sagan, and Valentino (2022) surveyed not only samples of American citizens but also samples from the United Kingdom, France, and Israel — three officially nuclear states and one *de facto* nuclear state. The experimental nature of the survey lies in the presentation to each national panel of a (clearly fictional) newspaper article. The article describes a scenario in which a terrorist group based in Libya is preparing a chemical weapons attack against the four respondents' capitals (Washington, London, Paris, Tel Aviv)[5].

---

[4] Since the atomic bombings of Hiroshima and Nagasaki, the official justification for this dramatic decision has been based on the high estimated cost in human lives that the U.S. armed forces would have had to endure in the final phase of World War II, had they launched a ground invasion of Japan (Tannenwald, 1999). On the connected "taboo" of the immunity of non-combatants (lawed in the IV Geneva Convention, 1949) see Kinsella 2011.

Experimental surveys have prompted various methodological and substantive critiques, to which we refer the reader to two dedicated debates published in *International Studies Review* (Smetana and Wunderlich 2021) and *Security Studies* (Pelopidas and Egeland 2023). Here, we will address two of these critiques: (a) certain limitations in the inputs introduced into the surveys in order to make them "experimental"; and (b) the stark discrepancy in respondents' willingness to support nuclear weapons use — relatively high in experimental surveys and significantly lower in standard ones. We begin by stating that, although we consider many of the critical observations directed at proponents of experimental surveys to be well-founded, we do not dismiss the underlying goal of these researchers: to find new ways of assessing and deepening our understanding of citizens' attitudes toward the use of nuclear weapons.

The first critique concerns the stimulation of respondents through inputs such as the fictional news article they are invited to read, and the degree of autonomy their responses can truly reflect within a "lab" context — one that is even more overtly artificial than standard survey settings. Of course, the use of stimuli such as cue cards, photographs, vignettes, or the presentation of documentary or custom-prepared texts is neither unknown nor unusual in social science research. However, it should be noted that such surveys typically address topics that are far less emotionally charged than those presented in the "mock news story," which includes casualty estimates for different types of attacks as well as highly influential details, such as the "opinion" of military leaders[6].

On highly emotional topics such as those involving the antithetic couples of security/insecurity, aggression/defense (ultimately: life/death), it becomes difficult to control the affective dimension of participants involved in these experiments. In response to critics who argue that "pre-emptive scenarios" involving terrorists, North Korea, or Iran are "tough tests," Sagan and Valentino reply that they are, rather, "realistic tests." By introducing figures and circumstances related to plausible enemies of the United States, they are testing "the nuclear taboo in exactly the [political] conditions in which it might be needed to prevent nuclear use" (Sagan and Valentino 2021: 1093). Rejecting the notion that their respondents are "too emotionally invested," the authors argue that, on the contrary, they may not be emotionally invested enough (Dill, Sagan and Valentino 2023: 197). Indeed, the experiments are quite far from "the depth of urgency and emotion that many Americans felt in 1945" and constitute "a conservative test of the public's willingness to use nuclear weapons and target members of non-combatants in war" (Sagan and Valentino 2017: 44).

---

[5] The dilemma posed concerns the means that each government should employ to neutralize the threat: whether to strike the terrorists using relatively less effective conventional weapons (with a 45% chance of destroying the target) or to resort to more decisive nuclear weapons (with a 90% chance of success). This choice is presented under the condition that both options would cause the same number of casualties (2,700), to be weighed against the estimated 3,000 victims if the attack were to occur (Dill, Sagan, and Valentino 2022). The data show that "American attitudes toward nuclear weapons and on combatant immunity are not exceptional […] contrary to many scholars expectations, the publics of two other nuclear-armed democracies are equally willing (France) as or more willing (Israel) than Americans to use nuclear weapons and deliberately target civilians" (Press, Sagan and Valentino 2022: 28).

[6] See the four-column headline: "Terrorists Planning Chemical Weapons Attack on Washington, D.C. [Paris, London, Tel Aviv]. Joint Chiefs Say Nuclear Airstrike Doubles Chances of Destroying Terrorist Chemical Lab in Libya," Appendix 1: News Stories (U.S. versus) in Dill, Sagan, and Valentino (2022).

Undeniably, manifestations that are not entirely rational are far from marginal aspects of the human condition and cannot be ignored. They are, in fact, integral to the worldview internalized by respondents — especially in contexts such as those described in experimental scenarios, which approximate a highly emotional reality. As current events in this third decade of the 21st century show, far from being dismissed as irrelevant or discredited as relics of the past, irrational elements are increasingly emerging as central actors in both public expressions and the behavior of decision-makers. For some time now, even the natural sciences have moved toward accepting — something the social sciences were the first to discover — that scientific knowledge is necessarily reductionist, destined to carve out limited segments of reality. In fulfilling its purpose, it is bound to reduce complexity at the very moment it seeks to impose cognitive order on what Max Weber called the "senseless infinity" (Weber 1904). The effort of scientific knowledge to recreate *in vitro* conditions that may represent even a distant proxy of historical reality cannot be dismissed, provided that researchers remain aware of the approximate nature of their endeavor.

Even more challenging is the second critique. Benoît Pelopidas and Kjølv Egeland argue that the findings presented in *Kettles of Hawks* — the largest of the experimental surveys conducted by the group (Dill, Sagan, and Valentino 2022) — "contrast with several existing polls and surveys that indicate relatively strong public opposition to nuclear use" in France, the United Kingdom, and the United States itself (Pelopidas and Egeland 2023: 190).

In our view, the explanation for this clear discrepancy does not lie primarily in differences of style, wording, or scenarios — although these certainly exist and undoubtedly contribute to distinguishing standard surveys from experimental ones. Rather, *the discrepancy stems from the difference in the object being measured*. The entire contrast between standard and experimental surveys can be resolved if we acknowledge that both are cognitively directed at two distinct, albeit related, phenomena. This is made possible through the valuable recovery of the distinction between the two concepts of opinion and attitude. The overlap between these two basic modes of human cognition had already drawn the attention of one of sociology's foremost figures, Robert K. Merton. In his theoretical interpretation of the *American Soldier* research campaign, Merton warned of the risk that "basic conceptual differences [...] might be obscured by a single crudely defined concept," such as "the single blanket concept of 'attitude'" (Merton and Kitt 1950: 73). Unlike *opinion*, which captures aspects of reality at a high level of abstraction, *attitude* reflects more practical and immediate aspects. Aware of the limited influence that political opinions had on soldiers' performance of their duties, the researchers led by Samuel A. Stouffer focused instead on "sentiments and attitudes" (e.g., "the willingness for combat") (Merton and Kitt 1950: 45).

Although the prevailing trend over time has been, improperly, to use the terms "attitude" and "opinion" interchangeably, the conceptual difference between them becomes even clearer when viewed through disciplinary lenses: as Bergman (1998) observes, social psychologists have typically favored the concept of attitude, political scientists that of opinion, and social anthropologists that of values. The usefulness of employing the concept of attitude as a predominantly affective and emotional manifestation has been reaffirmed, among others, by Lewis Aiken (2002), as distinct from the concept of opinion, which expresses explicit and conscious positions — what we might cautiously call "rational."

Applied to our topic, the controversy between standard and experimental surveys can be reconciled by explicitly recognizing the distinct objects targeted by each research tool: on the one hand, standard surveys aim to capture a detached and "rational" representation (albeit limited by abstraction); on the other, experimental surveys aim to capture an involved and "emotional" representation (albeit shaped by the constraints of the laboratory setting).

*2. A Survey: Opposition to the Use of Nuclear Weapons Between "Deontologists" and "Consequentialists"*

On the topic of public opposition to — or acceptance of — the use of nuclear weapons, the distinction between those who reject *a priori* such a possibility (*deontologists*) and those who base their judgment on an *a posteriori* evaluation of the expected (positive or negative) consequences (*consequentialists*) proves especially useful (Dill, Sagan and Valentino 2022). The crucial role of consequences in political decision-making — and thus the centrality of accepting responsibility for them — is fundamental in classical liberal thought, and particularly in Weberian sociology.

Hence, the unavoidable dilemma that, according to Weber (1919), accompanies most decisions in the political sphere: between the ethics of conviction (based on adherence to values) and the ethics of responsibility (based on consideration of consequences).

To assess the nature and distribution of opinions (in the narrow sense described in the previous paragraph), we conducted a standard survey on a sample of 802 Italian citizens aged 18 and older, representative by gender, age, and geographic residence in a country like Italy, which (at least as of spring 2025) participates in the American extended deterrence framework.

Beginning with perceptions of whether nuclear weapons are distinct from conventional weapons, 81% of respondents consider them fundamentally different and believe their use is always wrong. In contrast, 10% view nuclear weapons as merely more powerful than conventional ones and find their use appropriate in certain cases. Finally, 9% responded with "don't know/no answer" (see fig. 1).

Fig. 1: When it comes to conventional and nuclear weapons, which of the following two statements do you agree with?

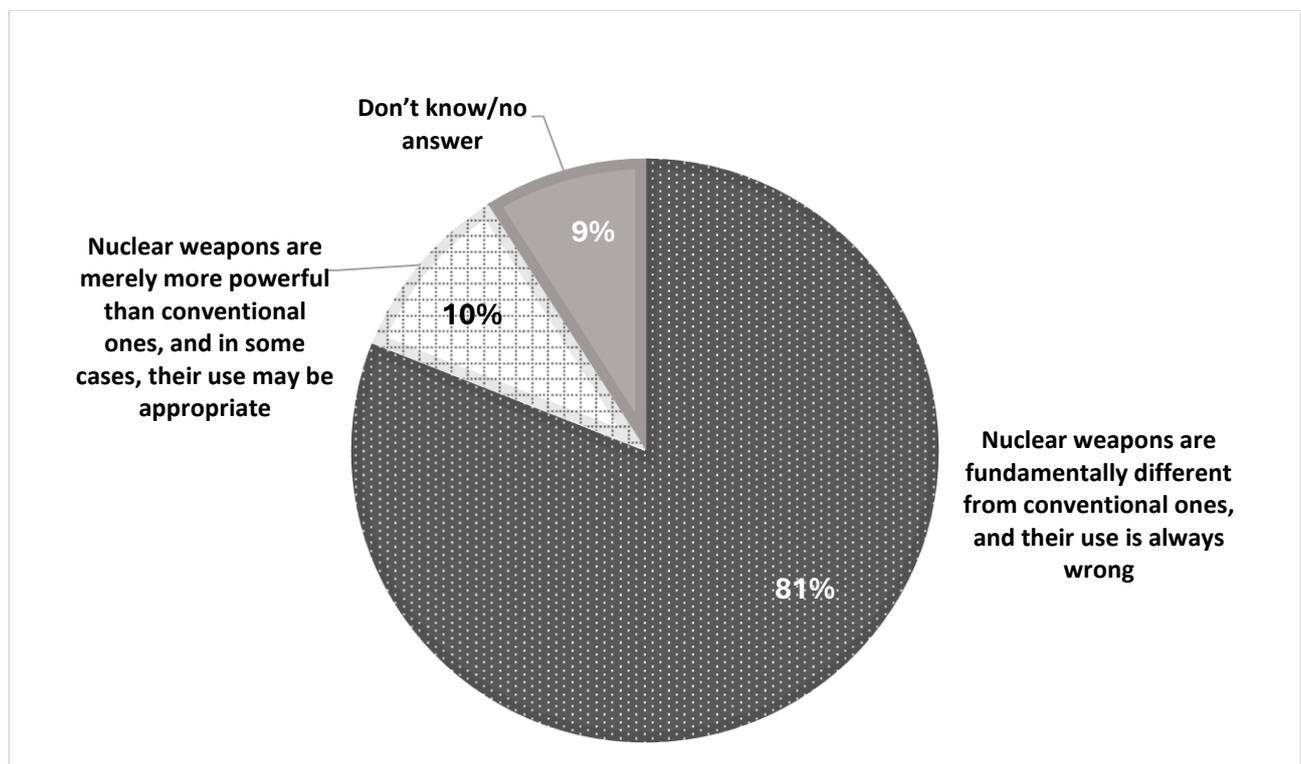

Archivio Disarmo, Difebarometro No. 11, February 2025. CAWI opinion survey on a sample of 802 respondents representative of the Italian population, January 2025.

Using a hierarchical binary logistic regression model, we first assessed the impact of sociodemographic variables on how respondents distinguish between conventional and nuclear weapons. Taking into account gender, age, education level, and political orientation, only age group and political orientation were found to be statistically significant. With respect to age, the younger the respondent, the more likely they are to consider nuclear weapons as merely more powerful than conventional ones and, in certain cases, their use as appropriate. As for political orientation, those identifying as right-wing or center-right are more inclined — compared to those on the left — to view the use of nuclear weapons as situationally justified.

We also asked respondents to answer two questions reflecting, respectively, an attitude and an opinion. The attitude involved a hypothetical situation: if one were the victim of a crime, would they turn to the police or take matters into their own hands, even using a weapon? The opinion concerns support or opposition to the introduction of the death penalty for the most serious crimes, such as mass murder or massacres. Two groups emerged: on one side, *reactive* respondents, who support harsh punishments and self-execution of justice; on the other, *adaptive* respondents, who support proportionate punishments administered by the State. The category of reactive respondents was subsequently added as an independent variable in the hierarchical binary logistic regression model.

Compared to non-reactive group, reactive respondents are statistically more inclined to believe that nuclear weapons are merely more powerful than conventional ones and that their use may be appropriate in certain cases. With the inclusion of the "reactive respondents" variable, age continues to show a statistically significant correlation with how respondents evaluate nuclear versus conventional weapons, while political orientation loses statistical significance. This change is due to the presence of a statistically significant association between political orientation and reactive respondents: those who identify as right-wing or center-right, rather than left-wing, are more likely to fall into the reactive group.

Subsequently, we asked the 81% of respondents (653 out of 804) who considered nuclear weapons fundamentally different from conventional ones and always wrong to specify the reasons for their position. The aim was to explore the broad range of responses which, as shown in the literature and in our previous research, reflect the variety of reasons underlying the nuclear taboo and the closely related issue of civilian immunity.

In order to systematize the range of responses into a conceptually structured framework, we identified three groups, arranged along the following axes: on the x-axis, the poles *A priori* / *A posteriori*; and on the y-axis, the poles Particularism / Universalism. Regarding the first axis, in the *A priori* category, opposition expressed by respondents is based on ethical and legal reasons; like the Kantian categorical imperative of "you must," a choice (in this case, the rejection of nuclear weapons) is made regardless of its costs or consequences, simply because it is morally right (hence the label "*Deontologists*" for those who hold this position)[7].

Conversely, in the *a posteriori* category, opposition is based pragmatically on a rational cost–benefit analysis focused on the expected consequences of a given decision (hence the term consequentialists for those who share this second position). Consequentialists, in turn, can be divided into two subgroups according to the Particularism/Universalism axis. While those who are unconditionally (*a priori*) opposed to nuclear weapons — presumably based on a universalistic principle — *a posteriori* opponents reject nuclear weapons for two distinct types of reasons. On one side are motivations that may be defined as *particularistic consequentialist*: they express opposition to nuclear weapons, but based on the harm that could befall the user, either through enemy retaliation or by creating a dangerous precedent that may backfire in the future. On the other side are motivations also centred on harmful consequences, but characterized by a broader perspective: harm done not only to oneself, but also to others. We define these as *universalistic consequentialist* motivations. They do not limit concern to foreseeable harm affecting oneself, one's family group, or nation, but extend it to the predictable harm suffered by all civilian populations (friendly, neutral, and even enemy), as well as by the environment.

---

[7] The ethical dilemma between a deontological choice and a consequentialist choice—illustrated by the well-known trolley problem—is extensively addressed in numerous studies of moral philosophy.

This conceptual classification of motivations for opposition to nuclear weapons use is presented in Fig. 2.

Fig. 2: Classification of Motivations for Rejecting the Use of Nuclear Weapons

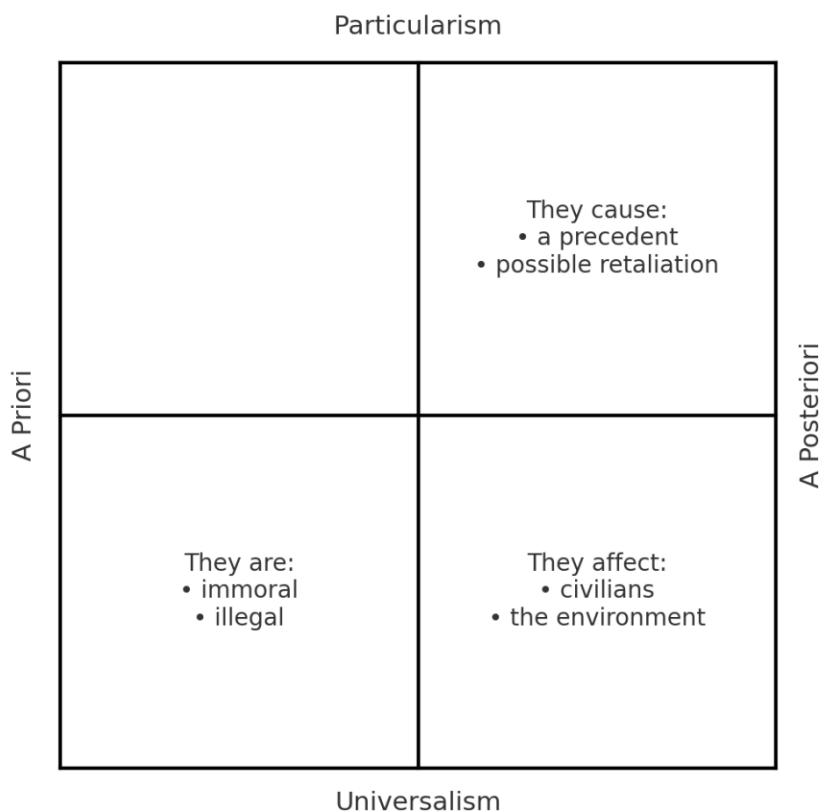

By applying the three categories of motivation to the 81% of respondents who consider nuclear weapons fundamentally different from conventional ones and believe their use is always wrong, we identified three corresponding groups. The largest of these (59%) consists of *universalistic consequentialists*. This group includes 51% who base their judgment on the serious and lasting harm nuclear weapons can inflict on innocent civilians, plus 8% who cite the environmental damage such weapons can cause.

In second place, at 23%, are the *particularistic consequentialists*: 15% are concerned about possible nuclear retaliation by one or more nuclear-armed states, and 8% are concerned about the precedent that might be set for states that possess — or could come to possess — nuclear weapons in the future.

Finally, 18% are deontologists (who, by definition, are universalists), believing nuclear weapons are always wrong due to their inherently immoral nature (12%) or their violation of international law (6%) (see Fig. 3).

Fig. 3: Breakdown of Motivations for Rejecting the Use of Nuclear Weapons

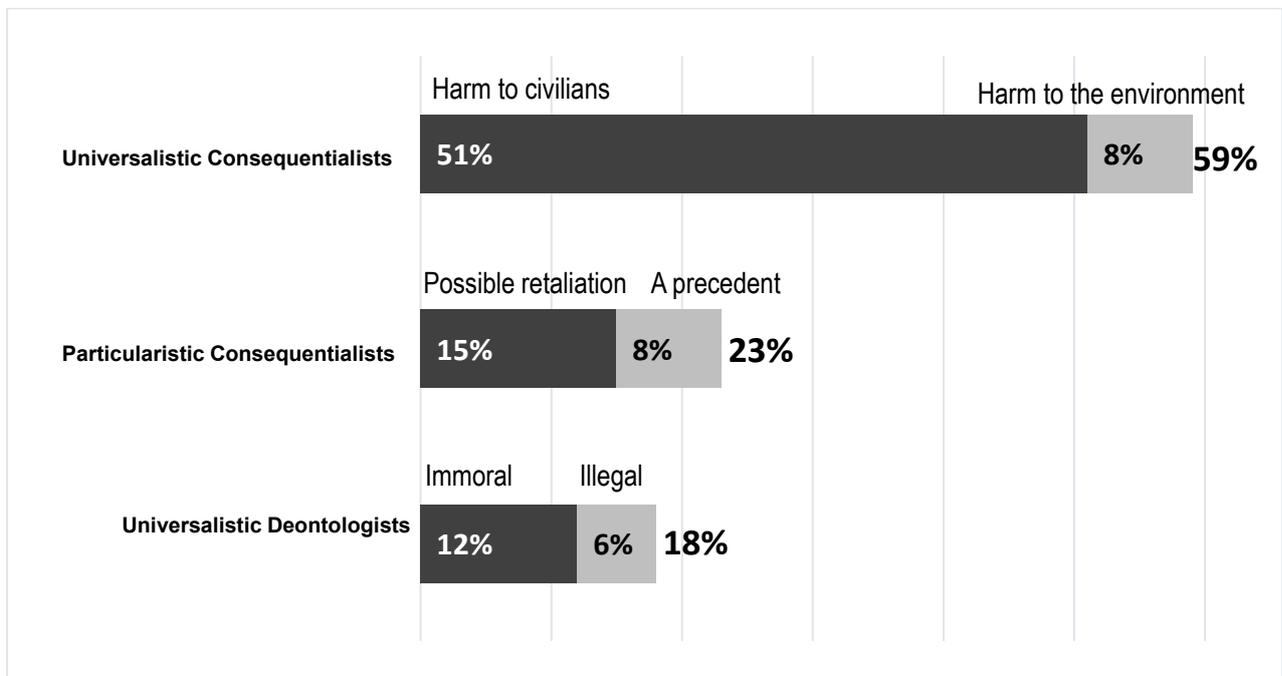

Source: Archivio Disarmo, Difebarometro No. 11, February 2025, sub-sample of 653 respondents who consider the use of nuclear weapons always wrong

Using a multinomial logistic regression model in which the most common category — *universalistic consequentialists* —was set as the reference group, we evaluated the impact of sociodemographic variables, as well as whether or not respondents belong to the category of reactive individuals, on the motivations underlying a negative assessment of the consequences of using nuclear weapons.

A first finding is the statistically significant impact of gender when comparing *universalistic consequentialists* with both *deontologists* and *particularistic consequentialists*. Specifically, the analysis shows that being male increases the likelihood of being a *deontologist* rather than a *universalistic consequentialist*. In the comparison between *particularistic* and *universalistic consequentialists*, both gender and reactivity play a role: being male increases the likelihood of adopting a particularistic stance, and being a reactive individual significantly increases the probability of being particularistic rather than universalistic. However, no statistically significant difference was found between *deontologists* and *universalistic consequentialists*.

Shifting from the general evaluation of nuclear versus conventional weapons to an analysis of the motivations for rejecting the use of nuclear weapons, we observe that gender — while not decisive in evaluating their use per se — has a statistically significant impact on why some respondents reject nuclear weapons. Women, more attuned to the practical consequences of their use rather than opposing them a priori, tend to adopt a universalistic consequentialist perspective more often than men, who are more likely to lean toward particularistic consequentialism.

Age, by contrast, does not appear to influence the reasons for rejecting nuclear weapons, but it does affect the general assessment of their use: as age increases, the likelihood of supporting the use of nuclear weapons decreases. Finally, being a reactive individual —defined by approval of the death penalty and armed self-defense — has an impact on both the evaluation of nuclear weapons use and the motivations associated with it. On the one hand, reactive individuals are more inclined to support the use of nuclear weapons in certain situations; on the other hand, among those who always oppose their use, reactivity is associated with particularistic consequentialist reasoning, based on fear of the potential harm their use could cause.

3. *Final Observations*

In a domain as highly complex as the one addressed in these notes, it is more appropriate to speak of approximations to a multidimensional reality than of definitive conclusions. Few issues present, as does that of nuclear armaments, such a complex intersection of variables — each one difficult to reduce to simple entities and, on the contrary, serving as a nexus for crucial physical, technological, psychological, social, economic, political, and ideological dimensions — just to name the main ones.

Each of these dimensions contributes to defining the issue of nuclear weapons and, more specifically for our purposes, their role in the eyes of public opinion. In our analysis, we have touched on the question of whether the term "taboo" is appropriate when referring to nuclear weapons. Should this term be considered excessive or not entirely suitable, it may be replaced with the less evocative and more descriptive term "aversion." The result does not alter the substance of the matter. In recent years, alongside the deterioration of international relations, there has been an effort — driven by politics and mainstream media — to "normalize" both the language and content associated with nuclear weapons. Despite undeniable progress in this direction, all national and cross-national public opinion surveys continue to show that this technology and strategic asset still inspire concern and hostility among the majority of the public in virtually every country in the world.

Whether this inhibition is vaguely sacred (a "taboo") or more secular (a "tradition") the use of nuclear weapons is still regarded, even by the elites of nuclear-armed countries, as a last resort: one that is heavily relied upon virtually (in terms of threats), but from which actual use has so far been abstained.

At the same time, no one can guarantee that what currently appears to be a reasonable expectation of no first use (supported by taboo or aversion — either way, by the "legacy of Hiroshima") will endure indefinitely. In the delicate balance whereby governing elites are, to some extent, influenced by public opinion, and citizens are influenced by nuclear weapons in the sense of opposing them, various and often contrasting factors interact in shaping the positions taken by both rulers and the ruled. On one side are rational factors (more or less aligned with a particular morality), and on the other, irrational factors (also converging, more or less, toward a different morality).

Our interpretation is that both types give rise to two distinct "products": reasoning manifests in opinions, while affectivity and emotion are expressed through attitudes. Since the human actor is an expression of both rational and affective-emotional components, both opinions and attitudes deserve recognition within individuals and groups, and both require scholarly investigation. It is therefore legitimate for different disciplinary and theoretical perspectives to choose which object to focus on. What matters most is that they do so with full awareness of their own epistemological and methodological strengths and limitations — and above all, that they succeed in bringing their findings to the attention of citizens and decision-makers, in pursuit of the shared objective of preventing nuclear war.